\documentclass[times, 10pt,twocolumn]{article}
\usepackage[compress]{cite}
\usepackage{latex8}
\usepackage{times}
\usepackage[figuresright]{rotating}
\usepackage{multirow}
\usepackage{booktabs}

\usepackage{amsmath}
\usepackage{subfig}
\usepackage{amsfonts}
\usepackage{amssymb}

\usepackage{graphicx}
\graphicspath{{C:/Users/Srikanth/Desktop/DCMLIpaper/Fig/}}
\usepackage{url}
\usepackage{framed}
\usepackage{setspace}
\begin{document}
	\title{
		Adaptive Nonlinear Control of High-Performance motors through \\ Multi-Level Inverters
	}
	
	\author{Srikanth Peetha, Michael L. McIntyre\\
		Electrical Energy Systems Research Group\\ University of Louisville, Louisville, KY, USA\\ s0peet01@louisville.edu, michael.mcintyre@louisville.edu \\
	}
	
	\maketitle
	\thispagestyle{empty}
	
	\begin{abstract}
		
		Many real-world systems are governed by the time-dependent, nonlinear differential equations. Dynamics of an electrical system are also best described using the very equations. Being one of the preferred machines when using advanced control algorithms for electric drives, DC motor has been selected for position tracking. Defining dynamic equations of the motor derived from the theory of nonlinear equations, aided our simulations by overcoming the time dependency that might not be addressed satisfactorily using regular control topologies. We incorporated integrator backstepping methodology along with a Multilevel Diode Clamped Inverter to develop full state feedback (FSFB) position tracking controller. The results thus obtained by using the integrator backstepping algorithm along with multilevel inverter were compared to the output of ideal case circuit that generated results based on integrator backstepping algorithm alone. Simulation results provide further validation with control topology being investigated for the ideal circuit and the inverter circuit.
	\end{abstract}

	\section{INTRODUCTION}
	Precision control is a vital aspect when dealing with high performance drive applications such as robotics, machine tools, rolling mills, and flying shears. They require high bandwidth drives which are insensitive to variations in operating conditions. Structural composition of these high-performance drives can be classified into three sections: an electrical subsystem, mechanical subsystem, and an electromechanical torque coupling of motor. The electrical subsystem dynamics are often neglected, for control simplification, as they are inherently faster than the subsequent mechanical subsystem dynamics\cite{c3}. An example of this control simplification is the common assumption that torque is the control input to the motor which is not the case in real-world systems. Even though the electrical dynamics are faster, they are defined by the nonlinear differential equations and are time dependent and hence cannot be neglected. Many real-world systems are found to be non-linear and time-dependent which are also governed by nonlinear differential equations. Hence a nonlinear control approach using \textit{Integrator Backstepping} \cite{c3} has been used to develop an adaptive control topology for achieving mechanical load position tacking objective. But attaining position tracking solely based on the control algorithm and backstepping is empirically not feasible. The reason for that is position tracking of load requires voltage control at high frequencies, and it should also address the harmonics and affects thus generated. Implementation of the nonlinear control algorithm can be facilitated with the help of a Diode Clamped Multilevel Inverter (DCMLI). Ability to incorporate pulse width modulation techniques along with advantages of producing high efficiency output, and the capability to perform the switching operations at the fundamental frequency motivated in selecting DCMLI \cite{c1} \cite{c2}. More over in multilevel inverters, as the number of inverter levels increases, the harmonic content decreases there by avoiding the need for filters \cite{c6}.
	
	The proceedings of our work are organized in the paper as follows. Section II shows the electrical and mechanical system models, and presents their mathematical representation. Working of a basic H-bridge inverter along with the operational standpoints of the diode clamped multilevel inverter and workings of the proposed control topology are described in section III. Development, design aspects of the control algorithm and analysis of their stability are addressed in Section IV. Finally, simulation results and conclusions are presented in sections V and VI respectively.
	
	\begin{figure*}
		\centering
		\includegraphics[width =0.8\textwidth]{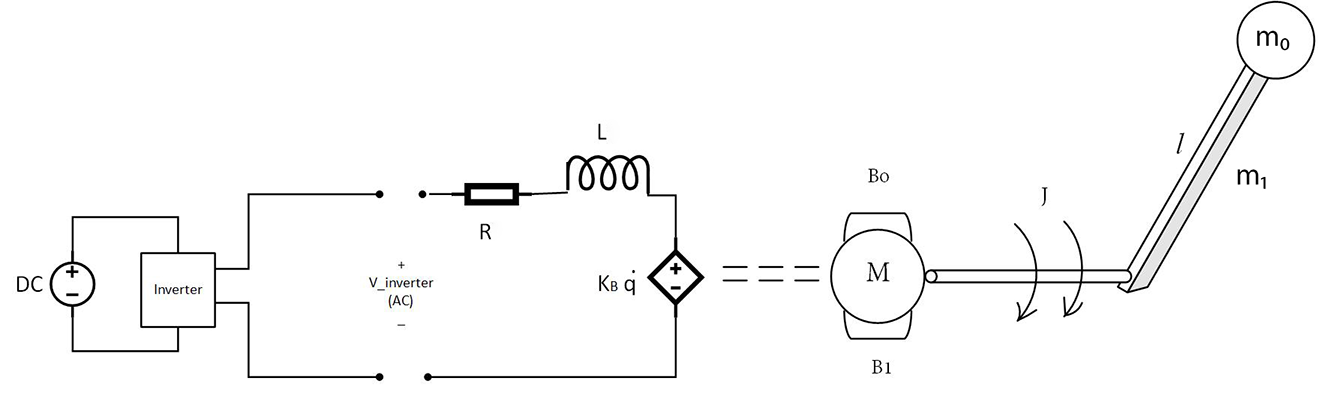}
		\captionsetup{justification=centering, font ={small}}
		\caption{Motor – Load Configuration}
		\label{Motor – Load Configuration}
	\end{figure*}

	\section{SYSTEM MODEL}
	Brushed DC motor with single joint mechanical load, and their corresponding dynamic equations are discussed in this section. Figure. \ref{Motor – Load Configuration} shows the equivalent circuit model of a Brushed DC motor along with the representation of mechanical arm load. The electrical subsystem has static and the dynamic components of the motor. Stator circuit of the DC motor has a series connection of resistor $(R)$ and an inductance $(L)$ while the rotor circuit contains brushes $(B_o, B_1)$ with an initial inertia of $J$, and a back-EMF coefficient of $K_B $ \cite{c3}.
	
	As shown in the figure. \ref{Motor – Load Configuration}, objective of the motor is to control the movement of mechanical load, of length $‘l’$ and a mass of ‘$m_0$’ and ‘$m_1$’, such that it follows a predetermined path ($q_d$). To make sure that load follows the required path, motor will be fed with an input voltage that is obtained by the controlled switching of a Voltage Source Inverter, about which we would discuss in the next section.
	
	The electrical subsystem dynamics of the system are expressed mathematically \cite{c3} as follows.
	\begin{equation}
	L\;\dot{I} = V-IR - \dot{q}K_B
	\end{equation}
	
	Where, $V$ is the input voltage applied to the motor that includes the amplitude modulation ratio $(u)$, given by the ratio of input and reference voltages \cite{c5}, $\dot{q}(t)$ is the angular load velocity, $K_B$ is the back-EMF coefficient, $I(t)$ is the rotor current and $\dot{I}(t)$ is the first derivative of rotor current.
	
	The mechanical subsystem dynamics of our position dependent load actuated by a brushed DC motor \cite{c3} are assumed to be in the form of
	
	\begin{equation}
	M\ddot{q} + B\dot{q} + N sin(q) = I
	\end{equation}
	
	Where $M, B, N$ are constant lumped inertia, friction coefficient, and lumped load term respectively. In this model, we assume that all the parameters $(M, B, N, L, R, and\; K_B$) to be constant and they are also defined to include the effects of torque coefficient constant which characterizes the electromechanical conversion of rotor current to torque \cite{c3}.

	\section{VOLTAGE SOURCE INVERTER}
	\subsection{H-Bridge inverter}
	
	An inverter is an array of power semiconductors and capacitor voltage sources, the output of which generates voltages with stepped waveforms. 
	
	\begin{figure}[h]
		\centering
		\includegraphics[width=\columnwidth]{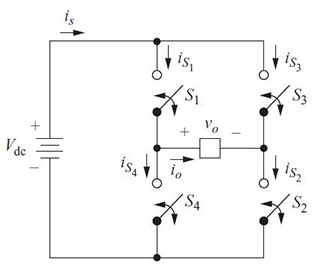}
		\captionsetup{justification=centering, font ={small}}
		\caption{H-bridge Inverter}
		\label{H-bridge Inverter}	
	\end{figure}
	
	Figure. \ref{H-bridge Inverter} shows the schematic of H-bridge inverter, a basic circuit that is used to convert DC to AC. In this circuit, each switching semiconductor device is replaced by an ideal switch. Table. \ref{Operation of H-bridge inverter} illustrates the operation of the inverter and the corresponding output voltages that are obtained because of the switching operations.

	The output voltage thus obtained because of the full-bridge inverter is a square wave with only three voltage levels ($+V_{dc}$, 0, $-V_{dc}$). To obtain an output that is similar to sinusoidal wave form, we must increase the number of voltage levels that are included in the output voltage. For that reason, we employ a multilevel inverter with diodes serving as voltage clampers.

	\begin{table}[h]
		\begin{center}	
			\captionsetup{justification=centering, font={small}}		
			\caption{Operation of H-bridge inverter}
			\label{Operation of H-bridge inverter}
			\begin{tabular}{|c|c|}
				\hline
				Switches Closed	& Output Voltage ($V_o$)\\
				\hline
				$S_1$ and $S_2$ & +$V_{dc}$\\
				$S_3$ and $S_4$ & -$V_{dc}$\\
				$S_1$ and $S_3$ & 0\\
				$S_2$ and $S_4$ & 0\\
				\hline
			\end{tabular}
		\end{center}
	\end{table}

	\subsection{Multilevel diode clamped Inverter}
\begin{figure}[h]
		\centering
		\includegraphics[width=\columnwidth]{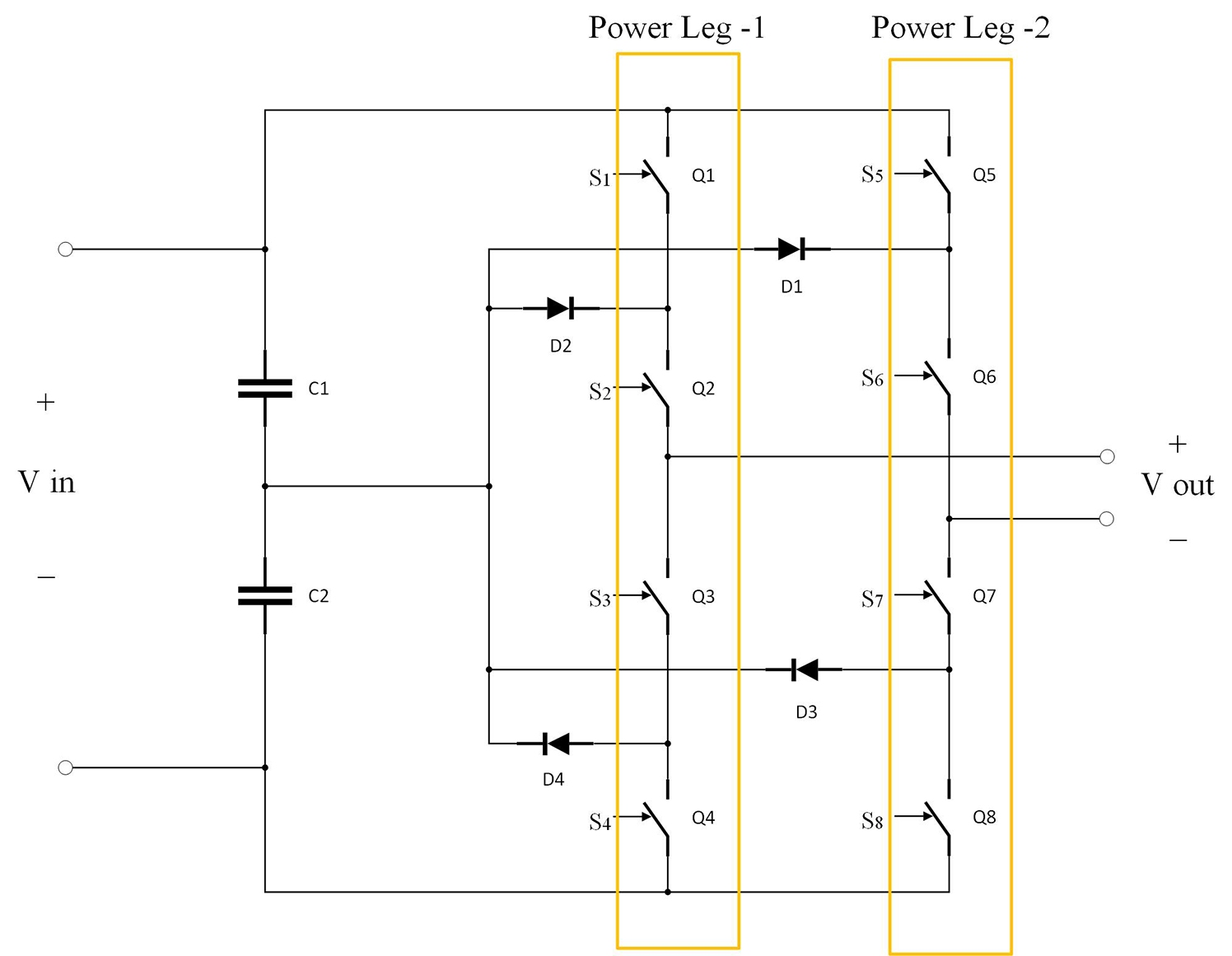}
		\captionsetup{justification=centering, font ={small}}
		\caption{Diode clamped Multilevel Inverter}
		\label{Diode clamped Multilevel Inverter}
	\end{figure}

	Figure. \ref{Diode clamped Multilevel Inverter} shows the circuit of a Diode Clamped Multilevel Inverter (DCMLI), also called as Neutral-Point Clamped (NPC) inverter. Unlike the full bridge inverter, DCMLI consists four switching elements in each power leg. DC source voltage that is fed into the circuit will split into five levels (\(+V_{dc}, \frac{+V_{dc}}{2}, 0, \frac{-V_{dc}}{2}, -V_{dc}\)) with the help of the PWM switching scheme. However, switching schemes of multilevel inverter tend to produce redundancies in voltage levels, meaning more than one switch combinations can generate the same output voltage level (line-to-line voltage). Care must be taken such that the redundancies are incorporated in the scheme and are addressed without unbalancing the voltage levels of capacitors.
	
	To control the switching operation of the inverter circuit effectively, a level shifted Sinusoidal Pulse Width Modulation (LSPWM) scheme is employed. The operation of LSPWM is explained in subsection $`C$'. The DC source voltage fed into the circuit is nearly constant and the amplitude of AC output voltage is controlled by adjusting the PWM ratio of the VSI. PMW ratio often varies with the delivered power, with higher ratio corresponding to the higher power. For each switching element, $Q_n$ shown in DCMLI circuit, a logical input signal, $S_n$, (whose value is either 0 or 1) is defined. Here $S_n$ = 0 denotes an ``off state'' and $S_n$ = 1 denotes ``on state''. For the proposed switching scheme, the switches are employed in complementary pairs \cite{c6} such that turning on one of the switches will exclude other switches from turning on i.e. $S_1$ = $\bar{S_3}$, $S_2$ = $\bar{S_4}$, $S_5$ = $\bar{S_7}$, $S_6$ = $\bar{S_8}$. Table. \ref{Switching Logic of the Inverter Circuit} shows the switching configuration used for the circuit.

	\begin{table}[h]
		\small
		\centering
		\captionsetup{justification=centering, font ={small}}	
		\caption{Switching Logic of the Inverter Circuit}
		\label{Switching Logic of the Inverter Circuit}
		\begin{tabular}{|c|c|c|c|c|c|}
			\hline
			&  &  &   &   &   \\
			State & $S_1$, $\bar{S_3}$ & $S_2$, $\bar{S_4}$ & $S_5$, $\bar{S_7}$ & $S_6$, $\bar{S_8}$ & $V_{out}$\\
			\hline
			0 & 0 & 0 & 1 & 1 & $-V_{dc}$ \\
			\hline
			1a & 0 & 1 & 1 & 1 & $-V_{dc}/2$ \\
			\hline
			1b & 0 & 0 & 0 & 1 & $-V_{dc}/2$ \\
			\hline
			2 & 0 & 1 & 0 & 1 & 0 \\
			\hline
			3a & 0 & 1 & 0 & 0 & $V_{dc}/2$ \\
			\hline
			3b & 1 & 1 & 0 & 1 & $V_{dc}/2$ \\
			\hline
			4 & 1 & 1 & 0 & 0 & $V_{dc}$ \\
			\hline
			
		\end{tabular}	
	\end{table}
	
	\begin{table}[h]
		\tiny
		\centering
		\captionsetup{justification=centering, font ={small}}	
		\caption{Simulation parameters \cite{c3}}
		\label{Simulation parameters}	
		\begin{tabular}{|c|c|c|c|}
			\hline
			&           &            &      \\
			Resistance (R) & 5$\Omega$ &    Load length $(l)$ & 0.305 [m]\\
			&           &            &      \\
			\hline
			&           &            &      \\
			Inductance (L) & 25$\ast$10$^{-3}$ [H] & Radius of load ($r_o$) & 0.023 [m]  \\
			&           &            &      \\
			
			\hline
			&           &            &      \\
			Constant controller Gain & 35 &  Back EMF ($K_B$) & 0.9 [N-m/A] \\
			($\alpha$) &  &  &\\
			&           &            &      \\
			
			\hline
			&           &            &      \\
			Mass of the link ($M_o$) &  0.434 [Kg]    &  Mass of the load ($M_1$) & 0.506 [Kg]\\
			&           &            &      \\
			
			\hline
			&           &            &      \\
			Controller gain($K_s$)& 10 &Controller gain($K_e$)&1\\
			&           &            &      \\
			
			\hline
			&           &            &      \\
			Adaptive gain matrix & diag &   Adaptive gain matrix &   \\
			$[\Gamma_\tau]$ & $\{0.01, 5, 5\}$   &  $[\Gamma_e]$ & 0.01*$I_{6X6}$ \\
			&           &            &      \\
			
			\hline
			                     &                     &                            &  \\
			                     &                     & Coefficient characterizing &  \\
			Graviatational force &                     & the electromechanical    & 0.9 N-m/A\\
			acting on load [G]             & 9.81 [Kg-m/$sec^2$] & torque conversion          & \\
			                     &                     & ($K_\tau$)                 &  \\
			                     &                     &                            &      \\
			\hline
			
		\end{tabular}
	\end{table}
	
	Where dig$\{.\}$ represents a diagonal matrix and $I_{6X6}$ is an identity matrix of size 6X6.
	
	The output voltage waveform obtained because of the inverter switching is similar to the reference waveform as shown in figure. \ref{PWM output}. The redundant states 1a, 1b, and 3a, 3b are implemented such that the states 1a, 1b are executed alternatively i.e. when a voltage level of $(-V_{dc}/2)$ is obtained the first time, the switching configuration corresponding to the state 1a is executed whereas the state 1b is executed the next time when $(-V_{dc}/2)$ is obtained. This way the scheme ensures that the bus splitting capacitors are evenly loaded.
	
	To perform the simulations, we need to derive the electrical and mechanical system parameters like $M, B, N$. The parameters  required are as given in the table. \ref{Simulation parameters}.

	\subsection{Level-shifted Sinusoidal PWM}
	
	Being one of the frequently employed techniques for traditional multilevel inverters, LSPWM technique serves as one of the path ways to decrease the total harmonic distortions of load current. Using this method, we can also control the switching operations thereby providing the flexibility to control the output voltage.

	\begin{figure}[h]
		\centering
		\includegraphics[width=8cm,height=6cm]{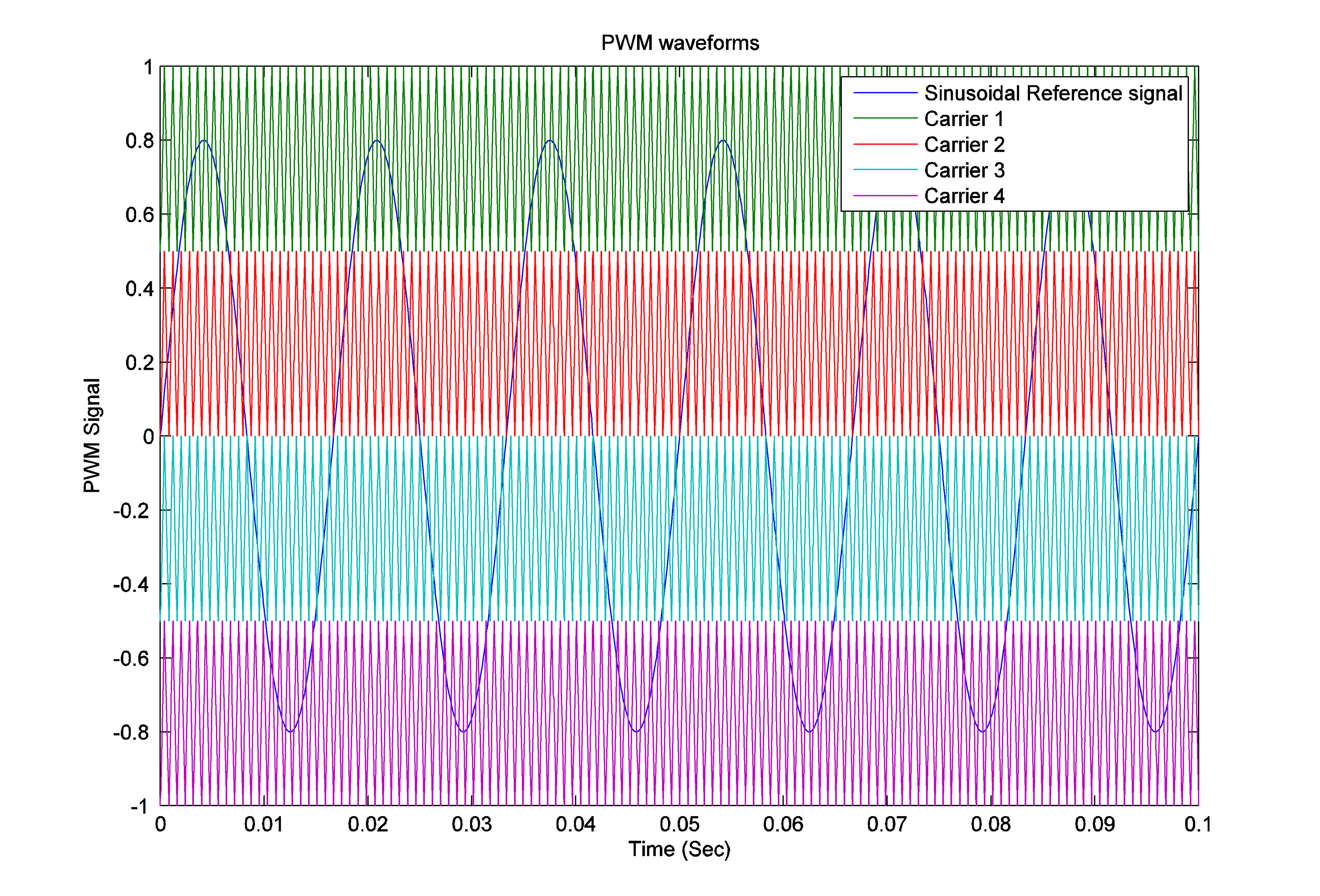}
		\captionsetup{justification=centering, font ={small}}
		\caption{Level shifted sinusoidal PWM technique}
	\end{figure}
	
	\begin{figure}[h!]
		\centering
		\includegraphics[width=\columnwidth]{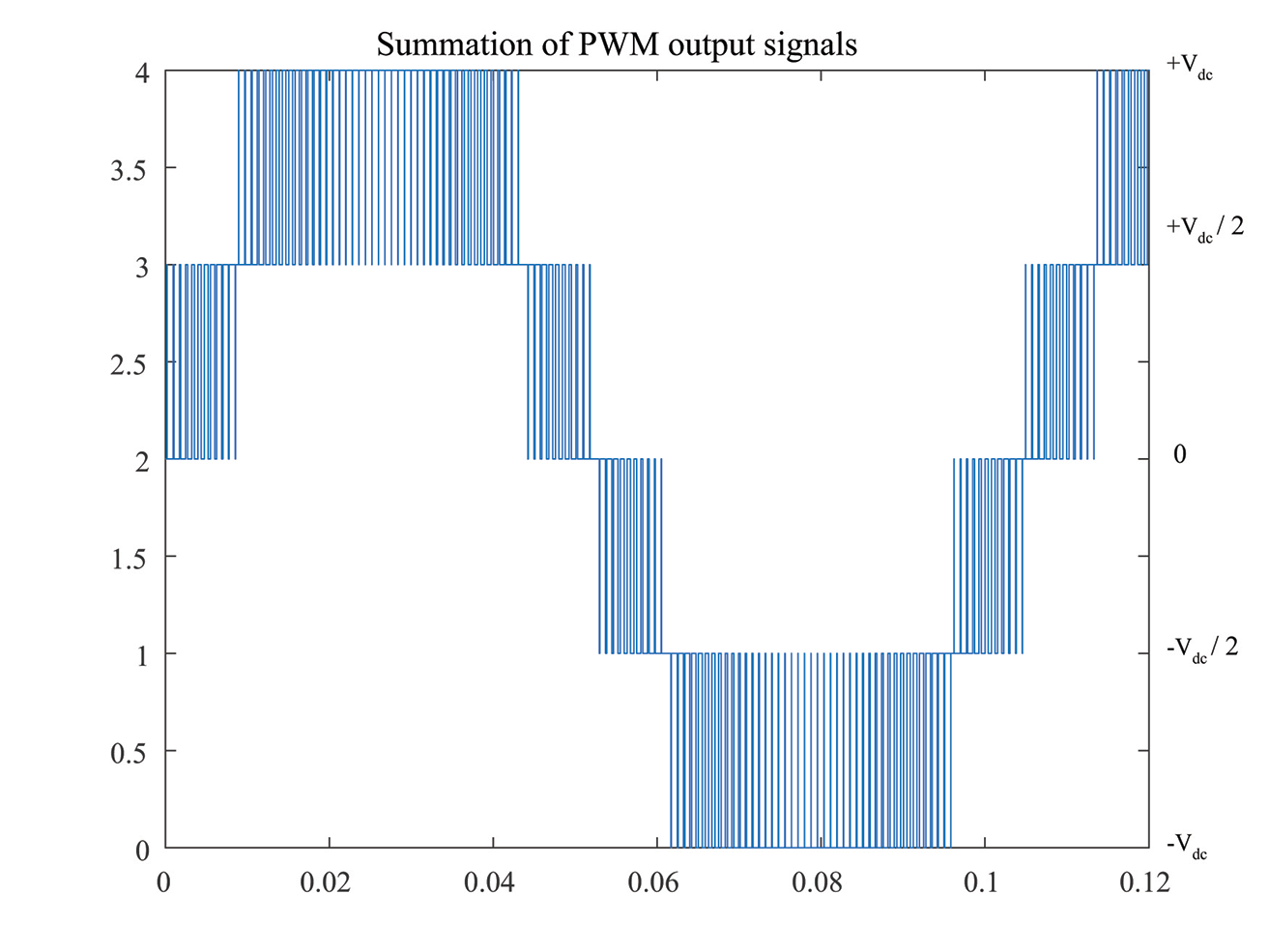}
		\captionsetup{justification=centering, font ={small}}
		\caption{PWM output}
		\label{PWM output}
	\end{figure}

	\begin{table}[h]
		\begin{center}
			\captionsetup{justification=centering, font={small}}		
			\caption{Operating principle of PWM scheme$\;$\cite{c1}}
			\label{table: Operating principle of PWM scheme}
			\begin{tabular}{|c|c|}
				\hline
				Comparision	& Output\\
				\hline
				Carrier signal $>$ Reference signal & 1\\
				\hline
				All other conditions & 0\\
				\hline
			\end{tabular}
		\end{center}
	\end{table}
	\setlength{\textfloatsep}{0.35cm}
	
	When an N-level inverter is employed, we need N-1 carrier signals \cite{c1} and all of them should have the same frequency, $f_c$, and same peak-to-peak amplitude. The zero reference will be set in the middle of the carrier signals. At every instant of time the level shifted carrier signals (triangular) are compared with a reference sinusoidal signal giving us the corresponding output voltage as show in table.		\ref{table: Operating principle of PWM scheme}. Later, at a given instant, all the output voltages thus obtained are added to obtain the final output voltage of the inverter (in the range of 0 to 4), giving five voltage levels as show in the figure.$\;$\ref{PWM output}.

	\section{Adaptive Controller Design}
	A nonlinear controller for position tracking is to be developed for the DC motor as described in the section above such that $q(t)$ $\rightarrow$ $q_d$(t) as t $\rightarrow$ $\infty$. In the condition where load path does not follow the desired path, we define load position tracking error $e(t)$ as.
	
	\begin{equation}
	e = q_d - q
	\end{equation}

	The desired position trajectory is selected as\cite{c3},
	
	\begin{equation}
	q_d(t) = \frac{\pi}{2} (1- e^{-0.1t^3}) \sin(\frac{8\pi}{5} t)
	\end{equation}
	
	To simplify the control algorithm formulation, we define filtered link tracking error, $r(t)$, as
	
	\begin{equation}
	r = \dot{e} +\alpha e
	\end{equation}
	
	Where $\alpha$ = positive controller gain\\
	
	Before proceeding any further, for mathematical calculations we make the following assumption: $q_d$(t), q(t) and all its derivatives are continuous and bounded functions of time, resulting in $e$ and $r$ to be bounded.
	
	The filtered tracking term, $r$, allows us to analyze the second order system as if it is a first order. To form the open loop filtered tracking error system, we differentiate the equation (4) with respect to time, we get
	
	\begin{center}
		$\dot{r} = \ddot{e} +\alpha \dot{e}$
	\end{center}
	
	Substituting (3) in the above equation, we get
	\begin{equation}
	\dot{r} = \ddot{q_d} - \ddot{q} +\alpha \dot{e}
	\end{equation}

	\begin{center}
		$(6) \ast M \rightarrow  M\dot{r} = M( \ddot{q_d} +\alpha \dot{e} ) +\ddot{q}M$
	\end{center}
	
	From equation (2) we have,
	\begin{center}
		$-M\ddot{q} = B\dot{q} + N \sin(q) - I$
	\end{center}
	\begin{equation}
	M\dot{r} = M(\ddot{q_d} +\alpha \dot{e}) + B\dot{q} + N\sin(q) - I
	\end{equation}
	
	We can rewrite the equation (7) in matrix form as
	\begin{equation}
	M \dot{r} = W_\tau \theta_\tau - I
	\end{equation}
	where $W_\tau$ and $\theta_\tau$ are the Regression Matrix and Parameter vector respectively, given by the equations 9 and 10.
	\begin{equation}
	W_\tau =[ \ddot{q_d} +\alpha \dot{e}  \kern1pc  \dot{q} \kern1pc  \sin(q)]
	\end{equation}
	\begin{equation}
	\theta_\tau = [M \kern1pc B  \kern1pc   N ]^T
	\end{equation}
	
	The mechanical subsystem error dynamics lack a true current (torque) control input. For this reason, we shall add and subtract the desired current trajectory $I_d(t)$ to the right-hand side of equation (8)
	
	\begin{equation}
	M \dot{r} = W_\tau \theta_\tau - I_d + \eta_I
	\end{equation}
	\begin{equation}
	\eta_I = I_d - I
	\end{equation}
	
	The above procedure of adding and subtracting the control input $I_d(t)$ is referred as \textit{Integrator Backstepping}. $I_d(t)$ is designed such that it would provide a good position tracking for mechanical systems (assuming it could be applied to the load).
	
	Considering the parametric uncertainty, we now design adaptive desired current trajectory for the mechanical dynamics of equation (11). Select $I_d(t)$ as,
	\begin{equation}
	I_d = W_\tau \hat{\theta_\tau} + K_s r
	\end{equation}
	$\hat{\theta_\tau} \in \mathbb{R}^3 $ represents dynamic estimate for the unknown parameter vector $\theta_\tau$ defined in equation (10)
	
	The parameter estimate $\hat{\theta_\tau}(t)$ defined in equation (13) is updated online according to the following adaptation law.
	\begin{equation}
	\hat{\theta_\tau} = \int_0^t \Gamma_\tau \; W_\tau(\sigma) \; r(\sigma)  \; \mathrm{d}\sigma
	\end{equation}
	
	Where $\Gamma_\tau \in \mathbb{R}^{3X3}$ is a Diagonal adaptive gain matrix (positive constant \& definite)\\
	
	Define mismatch between $\hat{\theta_\tau}$  and $\theta_\tau$  as
	\begin{equation}
	\tilde{\theta_\tau} = \theta_\tau - \hat{\theta_\tau}
	\end{equation}
	
	Substituting (14) in (15) and differentiating it with time, we get
	
	\begin{center}
		$ \frac{\mathrm{d}}{\mathrm{d} x} \tilde{\theta_\tau} = \frac{\mathrm{d}}{\mathrm{d} x} \theta_\tau - \frac{\mathrm{d}}{\mathrm{d} x} \big[  \int_0^t \Gamma_\tau \; W_\tau(\sigma) \; r(\sigma)  \; \mathrm{d}\sigma  \big] $
	\end{center}
	
	\begin{equation}
	\dot{\tilde{\theta_\tau}} = -\Gamma_\tau \; W^T_\tau \; r
	\end{equation}
	
	Substituting (13) in (11), we get
	\begin{equation}
	M \dot{r} = W_\tau \tilde{\theta_\tau} - K_s r + \eta_I
	\end{equation}
	
	Time derivative of the equation (13) results in
	\begin{equation}
	\dot{I_d} = \dot{W_\tau} \hat{\theta_\tau} + W_\tau \dot{\hat{\theta_\tau}} + K_s \dot{r}
	\end{equation}
	
	Substituting (9), (10) and (14) in the above equation,
	\begin{center}
		$\dot{I_d} = \hat{M} \big[  \dddot{q_d} + \alpha( \ddot{q_d}-\ddot{q} ) \big] + \hat{B} \;\; \ddot{q}$
	\end{center}
	\begin{center}
		$+ \hat{N} \;\; \dot{q} \;\; \cos(q)  +  W_\tau \;\; \Gamma_\tau \;\; W_\tau^T \;\; r $
	\end{center}
	\begin{equation}
	+ K_s \; (\ddot{q_d}-\ddot{q} + \alpha \; \dot{e})
	\end{equation}
	
	Where $\hat{M}, \hat{B}, \hat{N}$ denotes scalar components of the vector $\hat{\theta_\tau}$
	\begin{equation}
	\hat{\theta_\tau} = [\;\hat{M} \;\;\; \hat{B} \;\;\; \hat{N}\;]^T
	\end{equation}
	
	Equation (12) is given by
	\begin{center}
		$ \eta_I = I_d - I $
	\end{center}
	
	Differentiating with respect to time and after necessary algebra we get the linear parameterized open-loop model of the form,
	\begin{equation}
	L \; \dot{\eta_I} = W_1 \; \theta_1  - v
	\end{equation}

	The known regression matrix $W_1$ ($ q,\dot{q},I,\hat{\theta_\tau},t $) $\in \mathbb{R}^{1X6}$ and the unknown vector $\theta_1 \in \mathbb{R}^{6}$ are defined as follows,
	\begin{equation}
	\theta_1 = \big[\; \frac{L}{M} \;\;\; \frac{LB}{M} \;\;\; R \;\;\; K_B \;\;\; \frac{LB}{M} \;\;\; L \; \big]^T
	\end{equation}
	
	\begin{center}
		$W_1 =[ W_{11} \;\;\; W_{12} \;\;\; W_{13} \;\;\; W_{14} \;\;\; W_{15} \;\;\; W_{16} ]$ 
	\end{center}

	Elements of the regression matrix $ W_{1} $ are given by,
	\hangindent=0.5in $ W_{11} = \hat{B}I - K_s I - \alpha \hat{M} I $
	\newline
	$ W_{12} = K_s \dot{q} - \hat{B} \dot{q} - \alpha \hat{M} \dot{q} $
	\newline
	$ W_{13} =I $
	\newline
	$ W_{14} = \dot{q} $
	\newline
	$ W_{15} = K_s \sin(q) - \hat{B}\sin(q) + \alpha\hat{M} \sin(q) $
	\newline
	$ W_{16} = \hat{M}\;\; \dddot{q_d} + \alpha \; \hat{M} \;\; \ddot{q_d} + W_\tau \;\; \Gamma_\tau \;\; W_\tau^T \;\; r+ K_s \;\ddot{q_d} + K_s \; \alpha \; \dot{e} + \hat{N} \; \dot{q} \; \cos(q)
	$
	\newline
	
	The control input voltage $V$ is derived \cite{c3} from the equation below, 
	
	\begin{equation}
	V = W_1 \; \hat{\theta_1} \; + \; K_e \; \eta_I \; + r
	\end{equation}
	\begin{equation}
	Where,\;\; \hat{\theta_1} = \int_0^t \Gamma_e \; W_1 ^T(\sigma) \; \eta_I(\sigma)  \; \mathrm{d}\sigma
	\end{equation}
	
	$\Gamma_e \in \mathbb{R}^{6X6} = $ Adaptive gain diagonal matrix (positive definite).

	\subsection{Stability Analysis}
	We analyze stability of the given system to determine whether the given system is running towards stability or not without exploding out to infinity, which is highly undesired in daily practice. Our primary objective is position tracking i.e. to make $q(t) \rightarrow q_d (t)$. Initially a the Lyapunov candidate will be defined to prove the stability of the entire system. Later a few theorems to prove the nature of the equations will be defined along with corresponding mathematical proofs. For a complete analysis on the stability of the system, readers are directed to reference publication \cite{c3} where the required background has been defined.

	\section{Simulation Results}
	MATLAB/Simulink environment along with the PLECS Blockset module were used to construct and perform the simulations on our circuits. The results were used to further validate the control algorithm developed. The development of electromechanical circuits for both systems is facilitated by the equations derived in section IV. 
	
	\subsection{Results in ideal case}
	Model of the system is constructed using the MATLAB/Simulink environment per the dynamic equations (1) and (2).  The initial load position and the initial values of voltage, current are given as input to the electrical subsystem. Current output thus obtained from the electrical subsystem together with the load position serve as input to the mechanical subsystem. The desired path and current wave forms, acting as reference points, are developed based on the equations defined in (4) and (13). 
	
	\begin{figure}[h]
		\centering
		\includegraphics[width=\columnwidth]{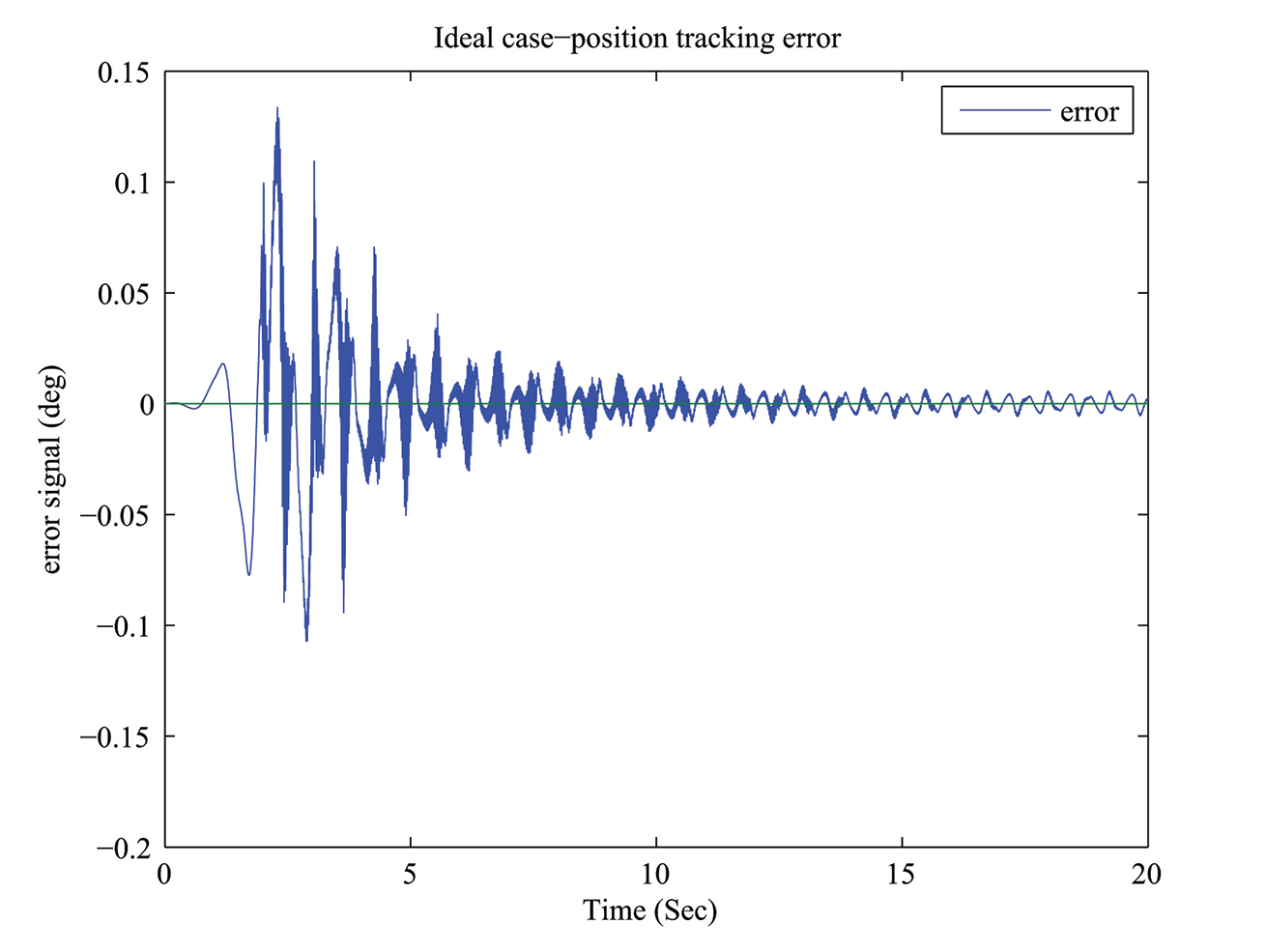}
		\captionsetup{justification=centering, font ={small}}
		\caption{Position tracking error signal of ideal case circuit}
		\label{Position tracking error signal of ideal case circuit}
	\end{figure}
	
	Whenever there is an error signal generated from the mismatch between the reference and the actual signals, it is redirected to the control algorithm which updates the value of output voltage thereby minimizing the error values. This voltage is then passed on directly to the motor, serving as input to motor. Because of the updated input, motor current there by the torque thus generated are varied, manipulating the position of the mechanical load. Figure. \ref{Position tracking error signal of ideal case circuit} shows the position tracking error signal obtained from the ideal circuit.

	\subsection{Results of Diode Clamped Inverter}
	\begin{figure}[h]
		\centering
		\includegraphics[width=\columnwidth]{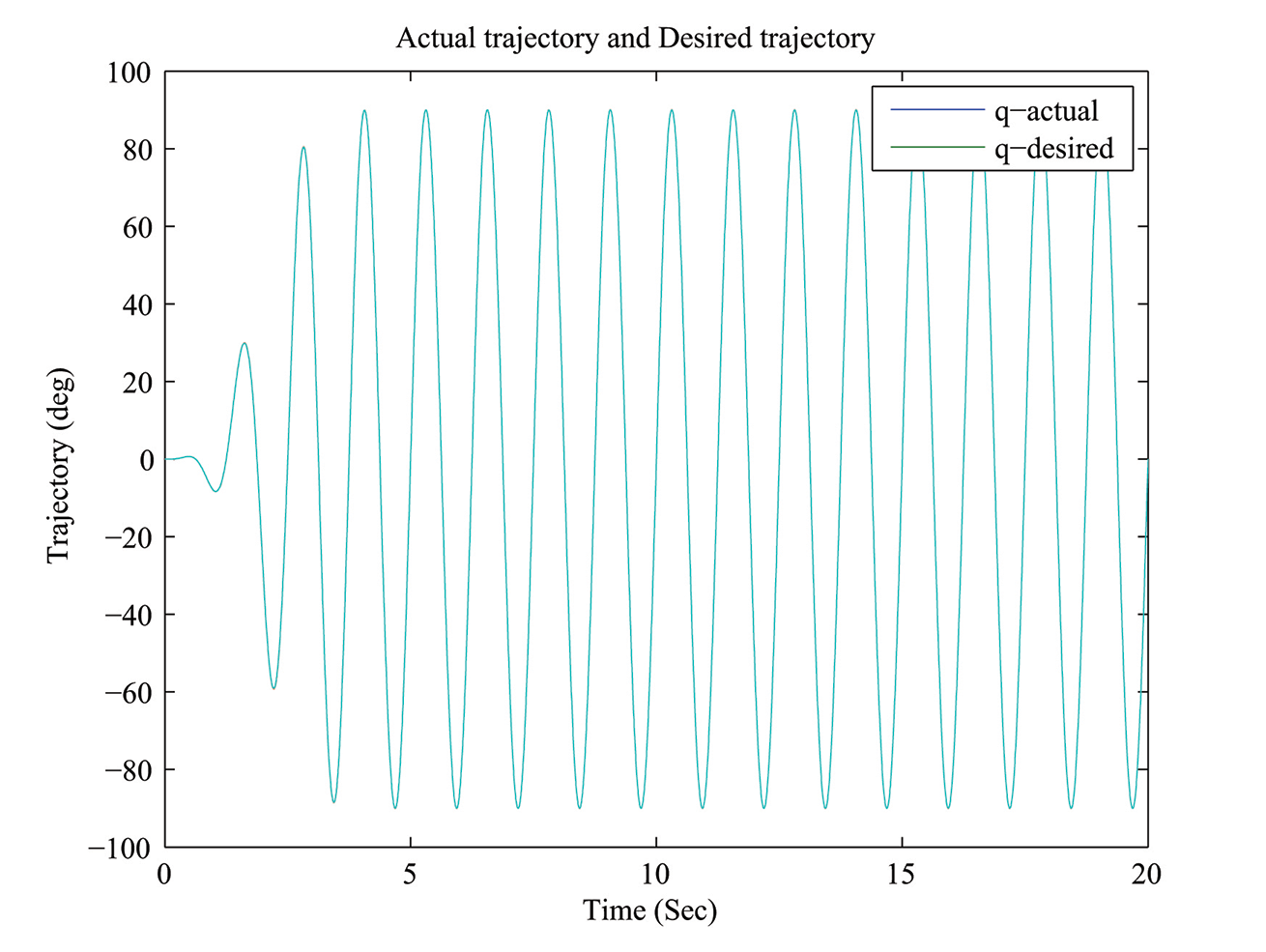}
		\captionsetup{justification=centering, font ={small}}
		\caption{Trajectories of actual path and the desired paths}
		\label{Trajectories of actual path and the desired paths}
	\end{figure}
	
	While using the inverter, instead of directing the output of voltage control algorithm to motor, we diverted it to the inverter and because of inverter's switching operations, a near sinusoidal output voltage wave form is obtained.
	
	\begin{figure}[h!]
		\centering
		\includegraphics[width=8.5cm,height=6cm]{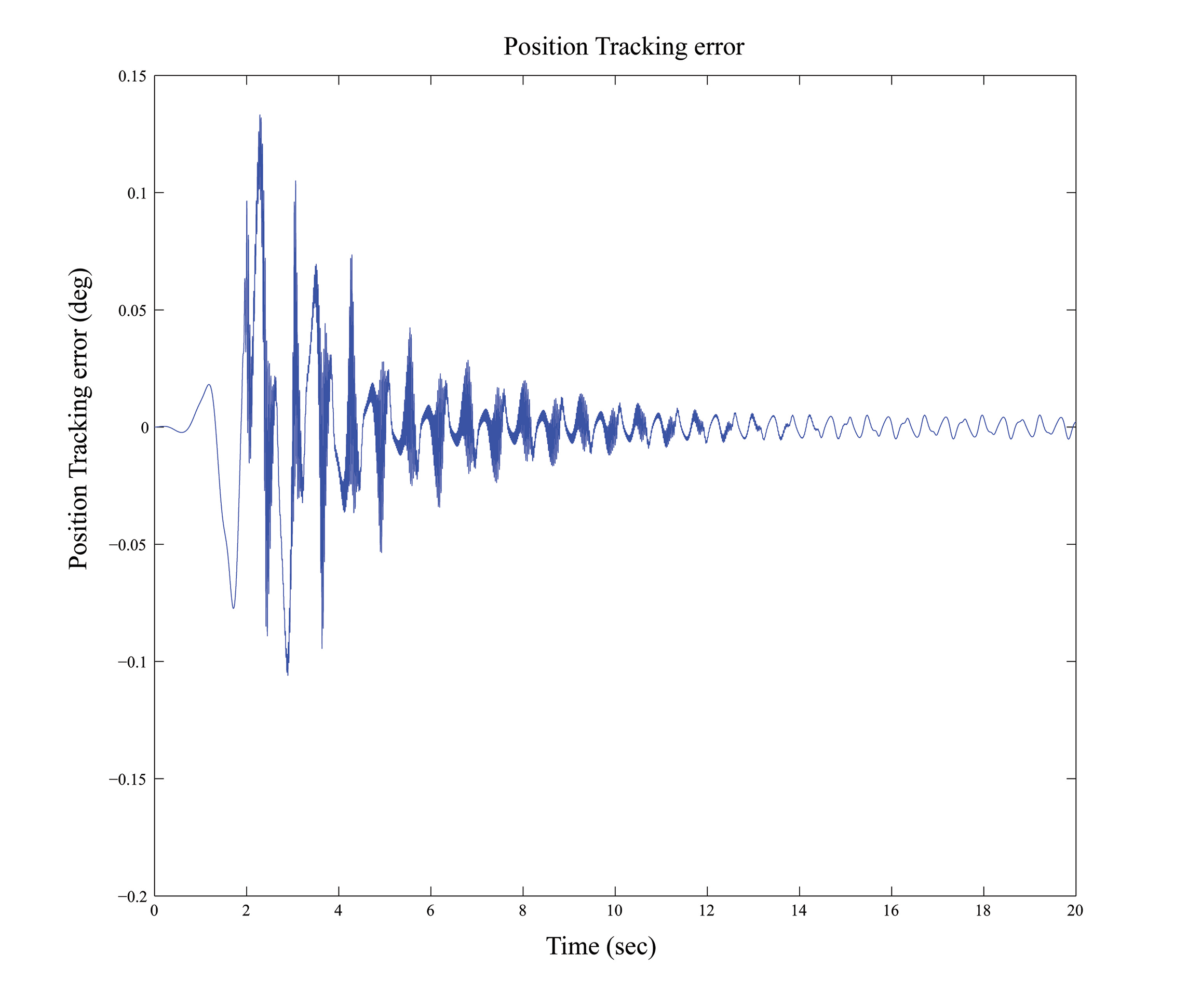}
		\captionsetup{justification=centering, font ={small}}
		\caption{Position Tracking error values obtained after including DCMLI}
		\label{ Tracking error values obtained after including DCMLI}
	\end{figure}
	
	\begin{figure*}[t]
		\centering
		\includegraphics[width=17cm,height=11cm]{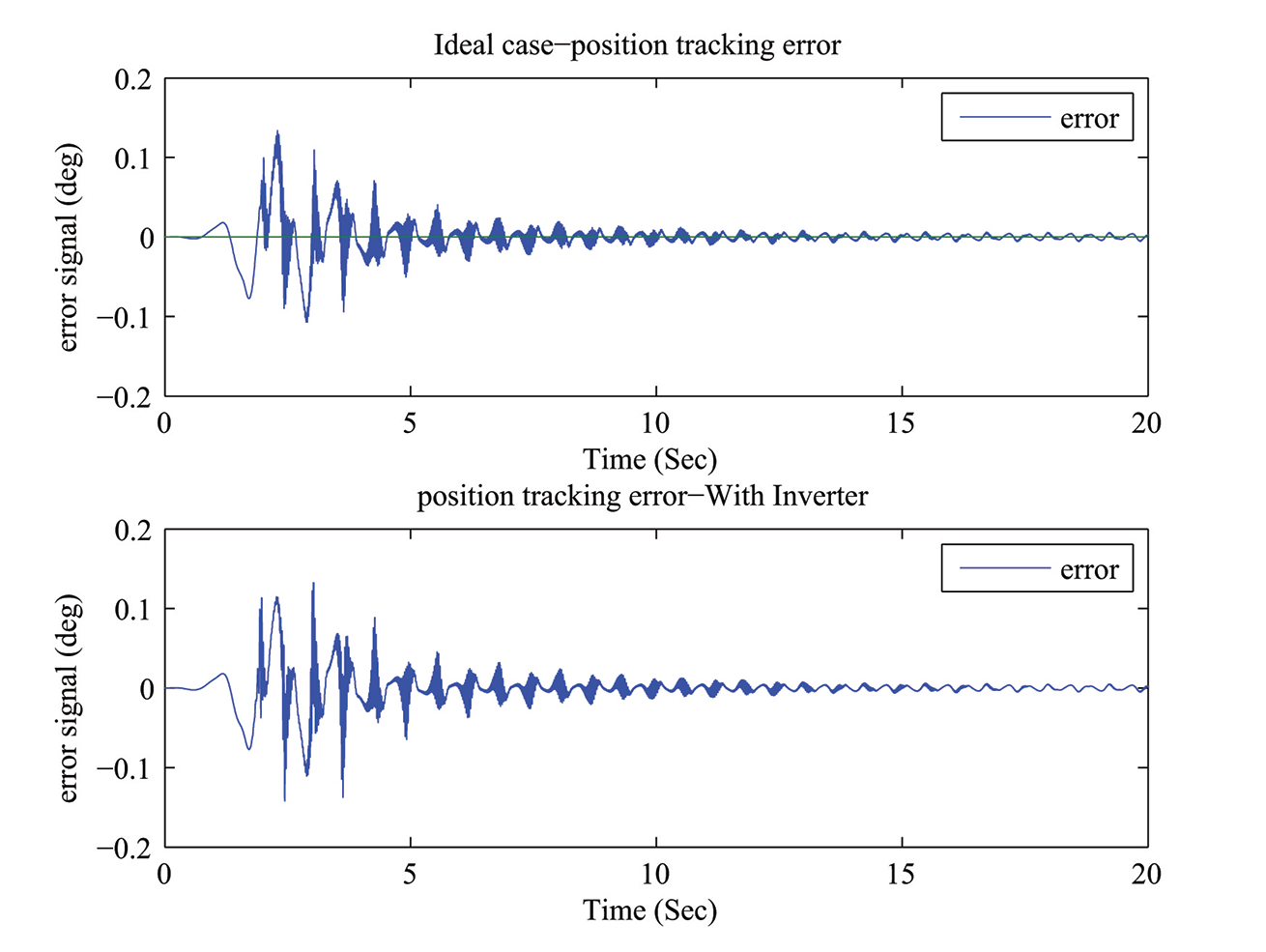}
		\captionsetup{justification=centering, font ={small}}
		\caption{ Comparison of error values -Ideal case circuit and inverter circuit}
		\label{ Comparison of error values -Ideal case circuit and inverter circuit}
	\end{figure*}
	
	\begin{figure}[h!]
		\centering
		\includegraphics[width=\columnwidth]{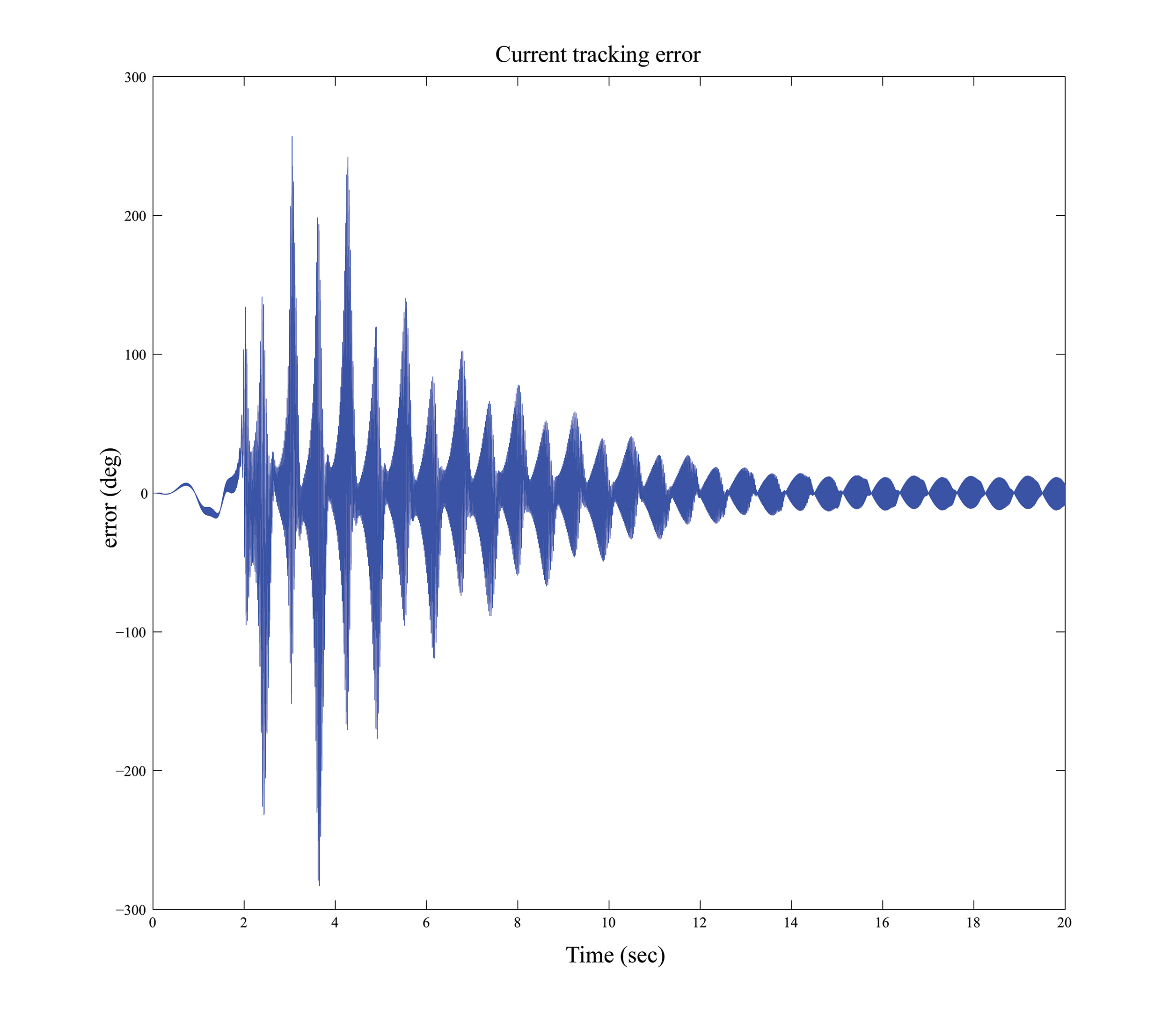}
		\captionsetup{justification=centering, font ={small}}
		\caption{ Current tracking error}
		\label{ Current tracking error}
	\end{figure}

\begin{figure}[h!]
		\centering
		\includegraphics[width=\columnwidth]{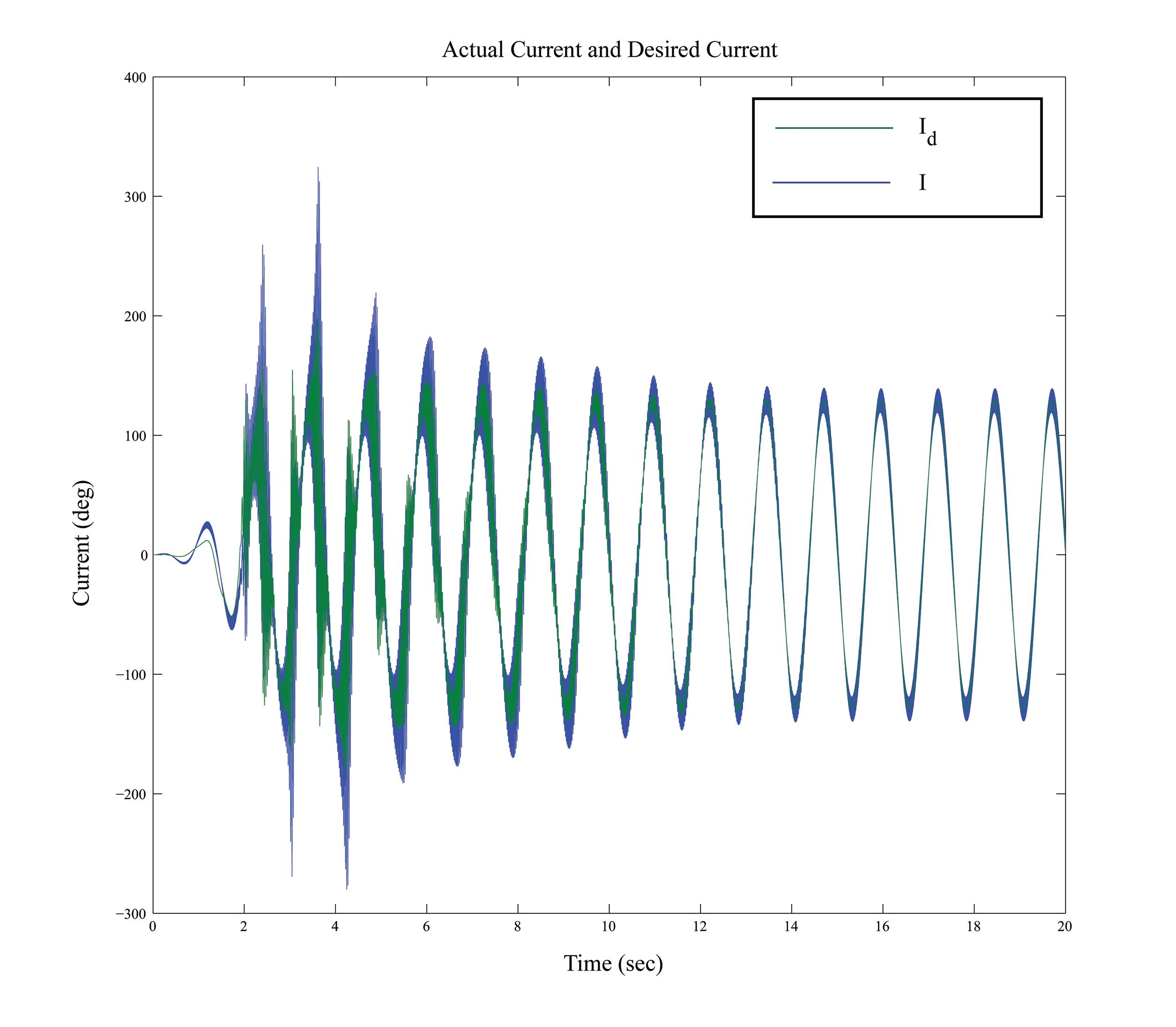}
		\captionsetup{justification=centering, font ={small}}
		\caption{ Current waveform tracking}
		\label{ Current waveform tracking}
	\end{figure}

	This inverter output is then given as input to the motor eliminating our primary obstacles, harmonics and common mode voltages. PLECS Blockset was used to model the circuit of multilevel diode clamped inverter as described in Section III. The output thus generated by the inverter is now given as input to the motor circuit and the results obtained are as shown in the figures \ref{Trajectories of actual path and the desired paths} through \ref{ Current waveform tracking}.
	
	Graph in the figure. \ref{Trajectories of actual path and the desired paths} shows combined waveforms of actual path $q_d(t)$ and the path followed by the motor $q(t)$. Even though both the desired and actual paths seem to be overlapping in the figure. \ref{Trajectories of actual path and the desired paths} there are a few error values because of position tracking algorithm, seen in figure. \ref{ Tracking error values obtained after including DCMLI}. But the error signal is gradually reduced to a near zero value proving the working capability of inverter and control algorithm developed. Moreover, from the plot shown in figure. \ref{ Comparison of error values -Ideal case circuit and inverter circuit} we can see that the error values generated by the simulation of DCMLI are as low as the error values of the ideal case circuit. From the figures \ref{Trajectories of actual path and the desired paths}, \ref{ Comparison of error values -Ideal case circuit and inverter circuit}, and \ref{ Tracking error values obtained after including DCMLI} we can conclude that $q(t) \rightarrow q_d(t) \;\; as \;\; t \rightarrow \infty $ signifying that position tracking objective is achieved by the DCMLI circuit as good as the ideal case circuit.
	
	The waveforms shown in figure. \ref{ Current waveform tracking} are the current signals generated from the filtered tracking error term defined in section IV. This graph shows successful implementation of current tracking objective. Like position tracking error signal, this wave form started at peak error values and stabilized gradually as time $(t) \rightarrow \infty $. The wave form for the error values of current tracking is plotted in figure \ref{ Current tracking error}. From figures \ref{ Current tracking error} and \ref{ Current waveform tracking}  it can be seen that the objective of current tracking is also achieved i.e.,$ \;\; I(t) \rightarrow I_d(t)\;\; as\;\; t \rightarrow \infty$.
	
	\section{Conclusion}
	A multilevel diode clamped inverter has been developed for a brushed DC motor. Use of level shifted sinusoidal PWM to govern the inverter, proved to be a reliable way to implement the position tracking objective by tracing down the error values to a minimum, also close to the ideal circuit. Integration of a multilevel inverter and integrator back stepping methodology eliminated the need for costly sensors and allows the controller to operate solely on the system dynamic equations. As explained in the chapter 2 of the cited book \cite{c1}, the stability of the equations considered and developed in this paper are proved. It has been proved via simulations that a diode clamped multilevel inverter can utilize the brushed dc motor effectively. Future work will include developing an experimental setup to further validate the proposed topology and extending our work towards induction motors, synchronous machines etc.

\end{document}